\documentstyle[aps,preprint]{revtex}
\begin{document}
\author{Hong Zhang\thanks{%
Author to whom correspondence should be addressed. }\thanks{%
Electronic address: hzhang@phys.hkbu.edu.hk; jinlinone@sina.com.cn}$^{1}$,
Bambi Hu$^{1,2}$, Gang Hu$^{1,3}$, Qi Ouyang$^{4}$, and J. Kurths$^{5}$}
\address{$^{1}$Department of Physics and Centre for Nonlinear Studies, Hong
Kong Baptist University, Hong Kong, China\\
$^{2}$Department of Physics and Texas Center for Superconductivity,
University of Houston, Houston, Texas 77204-5506\\
$^{3}$Department of Physics, Beijing Normal University, Beijing 100875, China%
\\
$^{4}$Department of Physics, Mesoscopic Physics Laboratory, Peking
University, Beijing 100871, China\\
$^{5}$Institute of Physics, University of Potsdam, PF 601553, 14415 Potsdam,
Germany}
\title{Turbulence control by developing a spiral wave with a periodic signal
injection in the complex Ginzburg-Landau equation}
\maketitle

\begin{abstract}
Turbulence control in the two-dimensional complex Ginzburg-Landau equation
is investigated. A new approach is proposed for the control purpose. In the
presence of a small spiral wave seed initiation, a fully developed
turbulence can be completely annihilated by injecting a single periodic
signal to a small fixed space area around the spiral wave tip. The control
is achieved in a parameter region where the spiral wave of the uncontrolled
system is absolutely unstable. The robustness, convenience and high control
efficiency of this method is emphasized, and the mechanism underlying these
practical advantages are intuitively understood.

PACS number(s): 82.40.Ck, 47.27.Rc, 92.60.Ek, 05.45.Gg
\end{abstract}

\newpage Turbulence exists popularly in nature, and it gives crucial
influence to the behavior of various systems. Thus, the investigation of
turbulence characteristics and turbulence control, as an extremely important
issue, has attracted continual attention for more than one century in a
large variety of fields of nature science, e.g. fluidic turbulence [1],
chemical turbulence [2], and electrical turbulence in cardiac muscle [3].
The model of the complex Ginzburg-Landau equation (CGLE) has been
extensively used for the study of turbulence and turbulence control problem.
The CGLE model reads$\quad \quad \quad \quad \quad \quad \quad $ 
\begin{equation}
\frac{\partial A}{\partial t}=A+(1+iC_{1})\nabla ^{2}A-(1+iC_{2})\left|
A\right| ^{2}A  \eqnum{1}
\end{equation}
which can be derived universally in the vicinity of a homogeneous Hopf
bifurcation in extended systems \cite{Aranson2002,Cross,Chate,Ouyang} with
the complex variable $A({\bf r},t)$ being the order parameter at the
bifurcation. Typical systems modeled by this equation include chemical
oscillations, transversely extended lasers and electrohydrodynamic
convection in liquid crystals.

In this paper we consider turbulence control by taking the CGLE as our
model. Since the pioneering work of OGY, chaos control techniques have been
well developed \cite{dfb,cfb,pp}, and the ideas of chaos control has been
applied extensively to spatiotemporal chaos control \cite%
{scc1,scc2,scc3,scc4,scc5,scc6,scc7}. Recently, turbulence control in
one-dimensional (1D) CGLE with gradient force has been investigated \cite%
{Xiao}. It was shown that for sufficiently strong gradient force one can
successfully control violent defect turbulence by injecting control signals
to few space points (even a single point) only. The reason for this high
efficiency is that the control effects can be propagated from the injected
space points to the points far away along the direction of the gradient
force, and then the influence of the control can reach large space areas not
directly controlled. In Ref. \cite{Aranson1}, Aranson {\it et al.} suggested
a method of turbulence control in the CGLE without gradient force by
developing a spiral wave with local feedback injection. The spiral wave
propagation from the central defect along the radial direction plays the
role of gradient force to transfer the control effect from the small
controlled tip region to the uncontrolled space region, and to greatly
enhance the control efficiency. Their method works when the spiral wave
state of the uncontrolled system is convectively unstable while fails when
it is absolutely unstable \cite{Aranson1}. In the present paper, we
investigate the turbulence control in 2D CGLE without gradient force also by
developing a regular spiral wave. Specifically, we focus on Eq. (1) with $%
{\bf r}=(x,y)$, $\nabla ^{2}=\partial ^{2}/\partial x^{2}+\partial
^{2}/\partial y^{2}$, and with no-flux boundary condition. The essential
advance of this work from Ref. \cite{Aranson1} is that we use a nonfeedback
control approach which can successfully suppress turbulence even when the
spiral wave of the uncontrolled system is absolutely unstable.

First, we fix $C_{1}=-1.4$, slowly increase $C_{2}$, and investigate the
behavior of the CGLE without any control. For a give pair of $C_{1}$ and $%
C_{2}$, the system has a spiral wave solution with a unique wave number $k$.
The spiral wave is stable for $C_{2}<C_{2}(C)\cong 0.37$ and convectively
unstable in the region $C_{2}(C)<C_{2}<C_{2}(A)\cong 0.79$ (see Fig. 1). A
careful adiabatic procedure allow (see Fig. 1), in principle, to avoid
breakup and to preserve on large spiral up to the absolute instability
boundary. Not too far from the absolute instability limit one is left with a
small spiral surrounded by turbulence whose radius does not depend on system
size and/or distance to the boundaries but vanishes as one approaches the
absolute instability threshold \cite{Aranson2,Aranson2002}. For $%
C_{2}>C_{2}(A)$, for the given wave number of the spiral wave of the CGLE
system, the perturbation growing rate becomes larger than the spiral wave
moving rate and the spiral wave of the uncontrolled system become absolute
unstable. For an arbitrary initial condition the system can quickly fall
into a turbulence state. In Figs. 1(a)-(d), we slowly changed $C_{2}$ from $%
0.77$ to $0.80$. In Fig. 1(a) we take $C_{2}=0.77$ in the convectively
unstable region and a stable spiral wave is observed. In Fig. 1(b), for $%
C_{2}=0.78$, near the absolutely unstable region, some turbulence appears
far from the spiral tip, while in Fig. 1(c) and 1(d), turbulence definitely
invade the spiral wave body and finally kill the spiral wave. In Fig. 1(e)
we show the distributions of stable, convectively unstable and absolutely
unstable region of the spiral wave solution.

In the convectively unstable regime of spiral waves, Aranson {\it et al.}
showed \cite{Aranson1} that they are able to suppress turbulence (bursts of
turbulence separated by the nucleation of well defined spirals) by local
feedback control in a small tip region, which can stabilize the tip defect
and develop an entire spiral wave to annihilating all other defects together
with turbulence. Their method works in the region of convective instability
and fails in the absolute instability $C_{2}>C_{2}(A)$, because the feedback
method is based on the existing spiral solution of the uncontrolled system,
which can never be stabilized by the local control near the tip area in the
absolutely unstable case \cite{Aranson1}. In the following we focus on
controlling turbulence in the region for $C_{2}>C_{2}(A)$, {\it i.e.},
controlling turbulence like Figs. 1(d) by applying convenient local
injection. In order to support and develop a spiral wave in the turbulent
surrounding of Fig. 1(d), we need a well behaved tip serving as the seed. In
experiments it is an easy matter to temporarily change the system parameters
in a small space area (e.g., in chemical reaction systems it can be done by
temporarily illuminating the given area by a light beam). With this change
we can generate a small spiral wave island in the violent turbulent sea. In
Fig. 2(a), we numerically generate this small spiral wave seed by changing $%
C_{2}$\ parameter to $C_{2}=0.5$\ and put a small spiral in the center $%
31\times 31$\ area for $t=-100$ t.u.. For $t\geq 0$, $C_{2}$\ returns back
to its normal value $0.8$. Then our task is: to support the spiral wave seed
of Fig. 2(a), and grow this seed into a large and entire spiral for killing
turbulence.

Let us start from Fig. 2(a) (remember the parameters of Fig. 2(a) are the
same as Fig. 1(d) in the whole space area, the small spiral seed serves only
as an initial condition). Without control, turbulence can easily invade the
center area of Fig. 2(a), and finally wipe out the spiral wave seed, leading
the system quickly from Fig. 2(a) to the fully developed turbulent state
Fig. 1(d), because the spiral wave of the system is absolutely unstable
without control. Our main idea of turbulence control is to inject a periodic
signal to a small fixed area around the initial spiral wave tip [{\it i.e.},
the center of Fig. 2(a)] to protect the spiral seed against the turbulence
invasion, and even to develop the spiral wave to annihilate turbulence in
the whole system. With control Eq. (1) is replaced by the following equation
for $t>0$%
\begin{equation}
\frac{\partial A}{\partial t}=A+(1+iC_{1})\nabla ^{2}A-(1+iC_{2})\left|
A\right| ^{2}A+\epsilon \delta _{i,\mu }\delta _{j,\nu }\exp (i\omega t) 
\eqnum{2}
\end{equation}
where $i,j$ are the integer numbers corresponding to the discretized $x$ and 
$y$ variables as $x_{i}=(i-1)$, $y_{i}=(j-1)$, respectively; $\mu ,\nu $ are
integer numbers. The control area is taken as a square in the space center
with $n\times n$ sites (for $n=1$, $\mu ,\nu =128$; for $n=2$, $\mu ,\nu
=127,128$; for $n=3$, $\mu ,\nu =127,128,129$ {\it etc.}). In Figs. 2(b)-(d)
we take $n=3$, $\epsilon =0.5$, and $\omega =1.2\omega _{0}$ ($\omega
_{0}=0.3762$ is about the angular frequency of the spiral of the CGLE model
(1) with $C_{1}=-1.4$, $C_{2}=0.5$) in Eq. (2). We find that with the signal
injection, the spiral seed can not only defeat the invasion of the
surrounding turbulence, but also develop to a large spiral wave body by
emitting waves into the turbulent region. In Fig. 2(d) the whole space is
firmly controlled by the spiral wave while the system parameters remain in
the absolutely unstable regime. The control efficiency in Fig. 2 is
surprisingly high and the approach is rather simple. We use only a single
signal injecting to $3\times 3$ space sites area extremely small in
comparison with the whole turbulent region of $256\times 256$ sites to turn
the violent turbulence to a perfect regular spiral wave. It is emphasized
that in our control process the development of the spiral wave in the
turbulent environment is not limited by the space area of Fig. 2. We have
tried to do the control same as Fig. 2 by increasing the system size to $%
512\times 512$, the control signals (the same as in Fig. 2) can continually
develop the spiral wave [with the seed same as Fig. 2(a)] until the original
turbulence is annihilated in the entire space. It is then interesting to
understand the mechanism underlying the above amusing control efficiency,
and the facts affecting the control results.

The existence of a small spiral wave seed is necessary. We have tried to
simulate the control equation (2) directly from the initial condition Fig.
2(a) without the spiral seed, turbulence could never be suppressed no matter
how we adjusted $\epsilon $, $\omega $, and $n$. The reason for this failure
is clear. There is no gradient force in Eq. (2), and usually the signal
injection can influence a very small region around the controlled region
only, and then no turbulence annihilation can be achieved \cite{Xiao}. With
a spiral wave seed, the situation is dramatically changed. Spiral waves have
a property to propagate the motion of tip region along the radial direction
convectively. With this convective propagation the spiral wave state itself
can transmit the effect of control signal from the tip region to far away
along the radial direction, behaving like a spiral-wave-induced effective
gradient force. Then, the mechanism of turbulence control of Eq. (2) [or
Fig. 2] in the presence of an initial spiral wave seed can be heuristically
understood. The injected signal plays role of stimulating the motion of the
spiral wave tip, and enhances its ability of generating and emitting waves.
The control effect can propagate along the radial direction, together with
these emitted waves. If the control signal can well match the tip motion,
the latter can be so well stimulated that the emitted waves can be strong
enough to stop the invasion of the surrounding turbulence and develop itself
to finally suppress turbulence.

Up to now we have not yet understood the exact meaning of match between the
injected signal and the spiral wave tip, but several facts related to this
matter can be intuitive explained, based on the numerical manifestations.
First, both too small and too large control forces are not favorable to the
successful control. In Fig. 3 we plot the asymptotic states of Eq. (2) for
different control strengths at $\omega =0.8\omega _{0}$ and $n=2$. The
optimal force intensity is around $\epsilon =1.5.$ It is easy to accept the
result of Fig. 3(a) because too weak control signal cannot inject enough
''energy'' to the spiral seed against the invasion of turbulence. The
understanding of the failure of large $\epsilon $ in Fig. 3(d) is a bit
nontrivial; the reason may be that too large forcing can modify the
structure of the spiral wave and reduce its ability emitting strong waves
for annihilating turbulence. Second, the frequency of the signal should be
properly chosen. In Fig. 4 we plot the system states at $\epsilon =0.5$, $%
n=5 $ and for different $\omega $'s. The successful control can be achieved
only for certain ''resonant'' frequency [Fig. 4(b)], too small [Fig. 4(a)]
and too large [Fig. 4(c) and 4(d)] frequencies give no good results. Here we
put quotation marks on the word ''resonant'' because the resonance is
nonlinear and depends on the forcing amplitude and the control area. And we
do not know precisely to which inner frequency the signal frequency is in
resonance. Third, the control area should not be too large for the control
purpose. In Fig. 5 we plot the full control region in the $\epsilon -n$
plane. For each pair of $\epsilon $ and $n$ the controllability is
determined by the optimal input frequency. For $n\geq 7$ we cannot achieve
full control whatever $\epsilon $ and $\omega $. This observation, which
looks a bit against the intuition, is actually not surprising. In our case
the periodic injection plays role in supporting the spiral tip. If the
signal is injected to too large area (note, $n=7$ is about a half of wave
length), the injection can cover both tip and arms of the spiral, and damage
the tip structure, and then the mechanism of turbulence control described
above no longer works.

In Fig. 2 the parameter $C_{2}$ is a bit above the absolute instability
boundary. We can control turbulence, by developing spiral wave, for the
parameter much deeply into the absolutely unstable region. For instance, in
Fig. 6 we do the same as Fig. 2 by taking $C_{2}=0.95$, $n=3$, $\epsilon
=1.18$, and $\omega =2.3\omega _{0}$. A successful control is obvious
achieved.

It is now interesting to ask why with the nonfeedback control we can
suppress turbulence in the parameter region where the feedback control
approach fails \cite{Aranson1}. The reason is clearly shown in Fig. 7 where
the solid line is the approximation of the separatrix between the absolutely
unstable region and the convectively unstable region in $k-C_{2}$ plane with 
$C_{1}=-1.4$ \cite{Aranson2,Aranson2002}. The circles in the figure show the
wave numbers of spiral wave of uncontrolled system in the convectively
unstable region. In this domain the feedback control can work well in
turbulence suppression by stabilizing a given defect, and developing an
entire spiral wave. Since the feedback control is based on these solutions,
the local control effect in the spiral wave center can function only
locally, and can not wipe out the turbulence in the region distant away from
the spiral wave center, {\it i.e.}, the small spiral wave can not grow to
eliminate the amplification of disturbances. With nonfeedback control we can
modify the wave number of the resulting spiral waves shifted by the external
periodic forcing from absolutely unstable regime to the corresponding
squares (within the convectively unstable region). This wave number
modification realizes the local stabilization of the spiral pattern and the
annihilation of the system turbulence.

In conclusion we have proposed a new and practical method of turbulence
control, based on the presence of an initial spiral wave seed. The approach
is convenient and effective because we need only to inject a single periodic
signal to a fixed and small space area. The method is effective even in a
region where the spiral wave of uncontrolled system is absolutely unstable.
We have tried different parameter regions for Eq. (2), and found similar
control results of Figs. 2 and 6. We expect that our method may be
effectively used in controlling turbulence in oscillatory and excitable
media which include a large variety of chemical and biological systems, but
it may not be applicable to the fluidic turbulence control with large
Reynolds number.

This work was supported in part by the grants from Hong Kong Research Grants
Council (RGC), the Hong Kong Baptist University Faculty Research Grant
(FRG), the National Nature Science Foundation of China, and the Nonlinear
Science Project of China. We would like to thank Lorenz Kramer, Wener Pesch,
Leihan Tang, Jinhua Xiao, Hongliu Yang, Zhuo Gao, Lei Gao and Quanlin Jie
for useful discussions.

\newpage

\section*{Figure Captions}

Fig. 1 $C_{1}=-1.4$ for all figures in this paper. (a)-(d) The asymptotic
spiral wave and turbulence solution of Eq. (1) by slowly increasing $C_{2}$
from $0.77$ to $0.80$. The spatial patterns are grey scale plot of $Re(A)$.
(a) $C_{2}=0.77$; (b) $C_{2}=0.78$; (c) $C_{2}=0.79$; (d) $C_{2}=0.80$. (e)
The distribution of stabilities of the spiral wave of Eq. (1). The
corresponding wave is stable for $C_{2}<C_{2}(C)\cong 0.37$ \cite{Aranson2};
convectively unstable in the region $C_{2}(C)<$ $C_{2}<C_{2}(A)\cong 0.79$
which is obtained from our numerical results of Fig. 1(a)-(d); and
absolutely unstable in the region $C_{2}>C_{2}(A)$. The local feedback
approach of Ref. \cite{Aranson1} can control turbulence in the region $%
C_{2}<C_{2}(A)$, our nonfeedback method can extend the effective control to $%
C_{2}>C_{2}(A)$. In all simulations we discretize the space variables to $%
256\times 256$ sites.

Fig. 2 (a) The spiral wave seed initiation for the control preparation. The
state (a) will be used as the initial state for all the following figures. $%
C_{2}=0.80$, $n=3$, $\epsilon =0.5$, $\omega =1.2\omega _{0}$, (b) $t=150$
t.u., (c) $t=450$ t.u., (d) $t=750$ t.u.

Fig. 3 The asymptotic states ($t=1350$ t.u.) of Eq. (2) for different
control strengths. $C_{2}=0.80$, $n=2$, $\omega =0.8\omega _{0}$. (a) $%
\epsilon =0.5$, (b) $\epsilon =1.5$, (c) $\epsilon =2.5$, (d) $\epsilon =3.5$%
.

Fig. 4 The same as Fig. 3 with parameters changed to $C_{2}=0.80$, $n=5$, $%
\epsilon =0.5$, and (a) $\omega =0.6\omega _{0}$, (b) $\omega =0.8\omega
_{0} $, (c) $\omega =1.0\omega _{0}$, (d) $\omega =1.2\omega _{0}$.

Fig. 5 The distributions of full control regions in the $n-\epsilon $ plane
with $\omega $ taken as the optimal frequency for each pair of $n$ and $%
\epsilon $. By full control we mean that turbulence can be completely
replaced by a perfect spiral wave in the asymptotic states ($t=1350$ t.u.),
like Fig. 2(d).

Fig. 6 The same as Fig. 2 with parameters changed to $C_{2}=0.95$, $n=3$, $%
\epsilon =1.18$, $\omega =2.3\omega _{0}$. (a) $t=800$ t.u., (b) $t=1200$
t.u., (c) $t=1600$ t.u., (d) $t=3000$ t.u.

Fig. 7 The wave numbers of the spiral waves of the uncontrolled [circles,
numerical results of Eq. (1)] and controlled [squares, numerical results of
Eq. (2)] CGLE systems vs $C_{2}$. The solid line is the approximation of the
separatrix between the absolutely unstable region and the convectively
unstable region in $k-C_{2}$ plane with $C_{1}=-1.4$ \cite%
{Aranson2,Aranson2002}. From the results of the uncontrolled spiral waves,
one can see that the wave numbers of spiral waves are in the absolutely
unstable regime when $C_{2}>C_{2}(A)$. The local periodic driving shifts the
spiral wave number from absolutely unstable regime to convectively unstable
regime (the corresponding squares). The control parameters of Eq. (2) are:
for $C_{2}=0.80$, $n=1$, $\epsilon =6.0$, $\omega =1.0\omega _{0}$; for $%
C_{2}=0.85$, $n=4$, $\epsilon =0.5$, $\omega =0.8\omega _{0}$; for $%
C_{2}=0.90$, $n=5$, $\epsilon =0.6$, $\omega =0.8\omega _{0}$; for $%
C_{2}=0.95$, $n=3$, $\epsilon =1.18$, $\omega =2.3\omega _{0}$.

\end{document}